\documentclass[epjc3]{svjour3}
\RequirePackage[T1]{fontenc}

\smartqed  
\RequirePackage[utf8]{inputenc}
\RequirePackage{graphicx}
\RequirePackage{mathptmx}      
\RequirePackage{flushend}
\RequirePackage{amsmath}
\RequirePackage{amssymb}
\RequirePackage{textcomp}
\RequirePackage{array,multirow}
\RequirePackage{slashed}
\RequirePackage{tabularx}
\RequirePackage{caption}
\RequirePackage{xcolor}
\RequirePackage{textcomp}
\RequirePackage{mathrsfs}
\RequirePackage[labelformat=parens]{subcaption}
\RequirePackage{booktabs} 
\usepackage[left=2.5cm,right=2.5cm,top=3cm,bottom=2.5cm]{geometry}
\RequirePackage[numbers,sort&compress]{natbib}
\RequirePackage[colorlinks,citecolor=blue,urlcolor=blue,linkcolor=blue]{hyperref}

\journalname{Eur. Phys. J. C}

\begin{document}
 \title{Study of neutrinoless double beta decay in the Standard Model extended with sterile neutrinos\thanksref{t1}}


\author{Debashree Priyadarsini Das\thanksref{e1,addr1}
        \and
        Sasmita Mishra\thanksref{e2,addr1} 
}

\thankstext[$\star$]{t1}{Thanks to the title}
\thankstext{e1}{e-mail: debashreepriyadarsini\_das@nitrkl.ac.in}
\thankstext{e2}{e-mail: mishras@nitrkl.ac.in}

\institute{Department of Physics and Astronomy, National Institute of Technology, Rourkela, 769008\label{addr1}
}

\date{Received: date / Accepted: date}

\maketitle
\begin{abstract}
We study a model where the Standard Model is augmented with 
three sterile neutrinos. By adopting a particular parameterization of a $(6\times6)$ unitary matrix - in this context, light neutrino masses being generated via a type-I seesaw mechanism -  we analytically derive the masses of the sterile states using an exact seesaw relation. The masses of the sterile states are derived in terms of the lightest mass of active neutrinos and
active-active and active-sterile mixing angles and phases; they can be all light, all heavy, or a mixture of light and heavy compared to the active states. This can be attributed to the interplay of the $CP$ violating (CPV) phases of the mixing matrix.  As both active and sterile states can mediate the neutrinoless double beta decay ($0\nu\beta\beta$) process, their contributions to the effective mass of the electron neutrino, $\lvert m_{ee}\lvert$, become a function of the mass of the lightest active state and active-active and active-sterile mixing angles and phases. We explore the parameter space of $\lvert m_{ee}\lvert$, 
keeping in mind, the present and future sensitivity of $0\nu\beta\beta$ decay searches. 
By making use of constraints from charged lepton flavor violating (cLFV) processes and non-unitarity, we explore the role of additional CPV phases and active-sterile mixing angle values. The numerical values thus obtained for $\lvert m_{ee}\rvert$ can vary from as low as $\mathcal{O}(10^{-4})$ to saturating the present experimental limit. We check the reliability of our result by calculating the branching ratio of $\mu \rightarrow e \gamma$, a prominent cLFV process, and non-unitarity in this framework.
\end{abstract}
\section{Introduction}
 \label{sec:intro}
The discovery of neutrino oscillation \cite{Jung:2001dh,SNO:2002tuh,K2K:2002icj,Hsu:2006gt} phenomena provides robust evidence in support of non-zero masses of neutrinos. This quantum mechanical phenomenon constrains beyond the Standard Model (BSM) physics but is unable to demystify the origin of neutrino mass or their absolute mass scale.  One of the instinctive ways to achieve the non-vanishing mass of neutrinos while preserving the Standard Model (SM) gauge symmetry is to extend the SM particle content through right-handed (RH) sterile neutrinos, often referred to as type-I seesaw models. In addition to providing a theoretical explanation of origin of the neutrino mass, sterile
neutrinos can help explain matter–antimatter asymmetry and dark matter. They can also be potential candidates contributing to the lepton number and flavor - violating process. In this work,
we study the role of sterile neutrinos, which give masses to the active neutrinos via a type-I
seesaw mechanism, in the lepton number-violating process of neutrinoless double beta decay, while
utilizing the constraints from non-unitarity and charged lepton flavor
violation (cLFV).

Various oscillation anomalies obtained through experiments like the Liquid Scintillator Neutrino Detector
(LSND) \cite{HARP-CDPGroup:2011hnr, LSND:2001aii} and MiniBooNE \cite{LSND:2001aii, MiniBooNE:2007uho, MiniBooNE:2010idf, Alvarez-Ruso:2021dna} allow the extension of SM particle content by one or more sterile neutrinos to accommodate the oscillation anomalies \cite{Kopp:2013vaa, Conrad:2012qt}.  While at least two sterile neutrinos must be added to the SM to explain the oscillation anomaly, there is an advantage to adding three as compared to two, as it makes more parameter space available for future searches of sterile neutrinos. For example, it was shown in Ref.\cite{Drewes:2021nqr} that the parameter space of active–
sterile mixing for which leptogenesis is attainable is larger in the case of three versus two extra sterile neutrinos.
Although sterile neutrinos have a perplexing mass scale and generation variation, their presence is strongly suggested for example in oscillation anomalies, cosmology, and collider physics. 

With the addition of $N$ number of sterile neutrinos, one can obtain a $(3+N)\times(3+N)$ neutrino mixing matrix $\mathcal{U}$ which includes extra active-sterile mixing angles and  CPV phases. In this study, we consider $N = 3$ and adopt a  parameterization developed in Refs. \cite{Xing:2011ur, Xing:2007zj} for the $6\times6$ unitary mixing matrix $\mathcal{U}$ in order to incorporate the active-sterile mixing parameters. In the intended $3+3$ model, the parameter space is spanned by active-active and active-sterile mixing angles, the CPV phases, and the lightest mass of the active neutrinos. However, the smallness of the non-unitary effects in the mixing matrix has to be kept in mind in order to comply with the neutrino oscillation data \cite{Xing:2007zj, Xing:2009in}, and this in turn implies small active-sterile mixing angles. 
Then, in the light of the exact seesaw relation, we obtain the analytical solutions for the masses of the three sterile states, to be determined in terms of the lightest mass of the active neutrinos and other mixing parameters.
 
The active and sterile neutrinos can occur with  Majorana mass terms and can trigger lepton number - violating (LNV) processes. In the context of LNV processes, $0\nu \beta \beta$ decay occurs as a leading phenomenon that can predict the absolute mass scale of neutrinos (particularly electron neutrino $\nu_e$) through its effective neutrino mass $\lvert m_{ee} \lvert$. However, the direct detection of $0\nu\beta\beta$ decay has not yet been achieved. Only experimental bounds on the half-life of $0\nu\beta\beta$ decays
(and hence $\lvert m_{ee}\rvert$) has been obtained for various isotopes from different collaborations. The most recent bounds on $\lvert m_{ee}\lvert$ obtained from  KamLAND-Zen \cite{KamLAND-Zen:2022tow} and GERDA \cite{GERDA:2020xhi} are ($0.036-0.156$ eV) and ($0.079-0.18$ eV), respectively. Also, the future ton-scale experiments CUPID \cite{CUPID:2019imh,Armengaud:2019loe}, LEGEND \cite{LEGEND:2017cdu}, and nEXO \cite{nEXO:2017nam,nEXO:2018ylp} aim for sensitivity of $\lvert m_{ee} \rvert$ as low as $0.01$ eV. Taking these 
experimental bounds on $\lvert m_{ee} \rvert$, we study the combined contributions of 
active and sterile neutrinos to the $0\nu\beta\beta$ decay process. 
As a consequence of an exact seesaw relation, $\lvert m_{ee}\lvert$ is a function
of the lightest mass of the active neutrinos and the mixing angle and phases of the $\mathcal{U}$ matrix. The numerical values for $\lvert m_{ee}\lvert$ are obtained
by varying the mixing angles and phases, keeping in mind the constraints
from neutrino oscillation data, non-unitarity, and the branching ratio (BR) of
cLFV processes.
Under these circumstances, the study of the effective mass $\lvert m_{ee}\lvert$ shows an affirmative result following the experimental bounds \cite{KamLAND-Zen:2022tow, GERDA:2020xhi}. Further, with inputs from present and future cosmological 
observations on sum of neutrino masses the combined contributions of active and sterile
neutrinos seem to provide more parameter space for the search of the $0\nu\beta\beta$ decay process.
Validation of the obtained values of mixing angles and CPV phases is reviewed with respect to a cLFV process of $\mu \rightarrow e \gamma$, which is a forbidden process within the SM and non-unitarity of the mixing matrix. As an example, we have elucidated certain sets of CPV phases and mixing angle values for which the BR ($\mu \rightarrow e \gamma$) is found to comply with the experimental bound of the Muon Electron Gamma (MEG) collaboration \cite{MEG:2011naj,baldini2018meg, baldini2021meg}.

The paper is organized as follows. In section (\ref{sec:type-I}), we introduce the $3+ 3$ model and the parameterization used for the $6\times 6$ mixing matrix. Also, by implementing the exact seesaw relation, we find the analytical expressions for the masses of the three sterile neutrinos in terms of   
lightest mass of the active neutrinos and mixing angles and phases of the $6\times 6$ mixing matrix. In section (\ref{sec:ovbb}), we discuss the combined contribution of active and sterile neutrinos to the $0\nu\beta\beta$ decay process. 
In section (\ref{sec:result}), we discuss the numerical results by comparing the combined
contribution with the standard active neutrino contribution in light of the present and future searches for the $0\nu\beta\beta$ decay process and cosmological observations. 
In section (\ref{sec:nonunitary-range}), we discuss the compatibility of the parameter space with the existing range of non-unitarity of lepton mixing. The following section (\ref{app:mu-e-gamma}) involves validating our results concerning the branching ratio of the cLFV process $\mu \rightarrow e \gamma$. We provide the conclusions of the work in section ({\ref{sec:conclusion}}). \ref{app:nme} is added to supplement the discussions in the previous sections.
\section{Type -I seesaw model with the inclusion of sterile neutrinos}
\label{sec:type-I}
With the addition of $N$ sterile neutrinos to the SM, the relevant part of the
Lagrangian is given by
\begin{equation}
-{\mathcal{L}} \supset i \overline{N_R}_{i}\gamma^\mu\partial_\mu {N_R}_{i} + \left( Y_{\alpha i} \overline{L}_\alpha \varPhi {N_R}_{i} 
+ \frac{{M_R}_i}{2} \overline{{N^c_R}_{i}}{N_R}_{i} + {\rm h.c.}\right),
\label{eq:lang-rhn}
\end{equation}
where $\varPhi$ and $L_\alpha = (e_{L\alpha}, \nu_\alpha)^T$, $(\alpha = e, \mu, \tau)$ are Higgs and lepton weak doublets, respectively. The Yukawa coupling matrix of the RH neutrinos ${N_R}_{i}$ are represented as $Y_{\alpha i}$. Here, we work in a basis in which the mass matrix of the charged leptons is diagonal. The Majorana mass matrix $M_R$ is taken to be real and positive. After electroweak symmetry breaking through the SM Higgs field acquiring a vacuum expectation value $\upsilon$, the mass matrix of the neutrinos on the basis of $(\nu_R^C~~ N_R)^T$ can be obtained as
\begin{equation}
 \mathcal{M} = 
 \begin{pmatrix}
  0 & m_D\\
  m_D^T & M_R
 \end{pmatrix}.
\label{eq:mass-matrix-tot}
\end{equation}

The mass matrix $\mathcal{M}$ is a $6 \times 6$ symmetric matrix which can be diagonalized by a unitary matrix of the same dimension.
Here, $m_D = Y \upsilon$ is the Dirac mass matrix. The magnitude of $M_R$ is experimentally unconstrained. 
In conventional type-I seesaw models, the mass scale of $M_R$ is
$\mathcal{O}(10^{15})$ GeV, requiring Yukawa coupling of the order 
$\mathcal{O}(1)$. However,
requiring a RH neutrino mass scale below the electroweak scale
but larger than the Dirac mass scale, the Yukawa coupling is fine-tuned with $\lesssim 10^{-12}$. Such fine-tuning can be realized as a
result of the breaking of some underlying symmetry such as the tiny breaking 
of lepton number symmetry \cite{Shaposhnikov:2006nn}.
Also, the light sterile neutrinos can be realized in models with popular $A_4$ symmetry \cite{Lindner:2010wr,CarcamoHernandez:2019kjy} or softly broken $L_e-L_\mu-L_\tau$ symmetry \cite{Barry:2011wb}. 
Also, motivated by the running of the Large Hadron Collider (LHC), more attention is being focused on searching for GeV - TeV scale heavy neutral leptons at colliders \cite{Atre:2009rg,Antusch:2016ejd,Abada:2022wvh}. They could be searched for in the decays of the mesons and $\tau$-lepton (a comprehensive review on collider, nuclear decay, extracted beamlines, and cosmological and astrophysical searches can be found in Refs.  \cite{Bolton:2019pcu,Abdullahi:2022jlv}).
The motivation for the low-scale seesaw also arises from the fact that the hierarchy problem could be avoided due to the contribution of superheavy RH neutrinos to the Higgs mass \cite{Haber:1993wf}. 

The  matrix diagonalizing $\mathcal{M}$ can be represented in $2\times 2$ matrix form, with each block being a $3 \times 3$ matrix as
\begin{equation}
 \mathcal{U} =
 \begin{pmatrix}
 V & R\\
  S & U
 \end{pmatrix},
\label{eq:curl-U}
\end{equation}

\begin{equation}
 \begin{pmatrix}
 V & R\\
  S & U 
 \end{pmatrix}^\dagger
\mathcal{M} \begin{pmatrix}
 V & R\\
  S & U 
 \end{pmatrix}^* = 
 \begin{pmatrix}
  \hat{m}_\nu & 0\\
  0 & \hat{M}
 \end{pmatrix},
 \label{eq:diagonalising}
\end{equation}
where $\hat{m}_\nu = \rm{Diag}(m_1, m_2, m_3)$ and 
$\hat{M} = \rm{Diag}(M_1,M_2, M_3)$ are the mass matrices of the active and sterile neutrinos 
in their respective mass bases. The diagonalization of the neutrino mass matrix can also be viewed as the flavor states of the active neutrinos expressed in terms of corresponding active and sterile mass eigenstates as

\begin{equation}
 \nu_{L\alpha} = V_{\alpha i} \nu_i
+ R_{\alpha i} N_{i}.
\label{eq:flavourstate}
\end{equation}
The standard charged-current interactions between the light neutrinos and charged leptons can now be represented as
\begin{equation}
 -{\mathcal{L}} = \frac{g}{\sqrt{2}}\overline{(e, \mu, \tau)_L} ~
 \gamma^\mu ~ \left[ V 
  \begin{pmatrix}
   \nu_1\\
   \nu_2\\
   \nu_3
  \end{pmatrix} + R 
 \begin{pmatrix}
  N_1\\
  N_2\\
  N_3
  \end{pmatrix} 
  \right] W^-_\mu + {\rm h.c.}.
  \label{eq:chrgd-crrnt}
\end{equation}
Here, the matrices $V$ and $R$ signify the active neutrino oscillation and strength of charged-current 
interactions of the sterile neutrinos, respectively. The unitarity condition of the matrix $\mathcal{U}$ given in Eq.(\ref{eq:curl-U}) leads to the condition 
\begin{equation}
 V V^\dagger + R R^\dagger = 1.
 \label{eq:unit-condition}
\end{equation}
The latter condition implies the non-unitarity of the matrix $V$ that can be quantified
from the measure of $|R R^\dagger|$. 
Using Eq.(\ref{eq:diagonalising}), we can establish a correlation between $V$ and $R$ through the exact seesaw relation \cite{Xing:2009ce} as 
\begin{equation}
V \hat{m}_\nu V^T + R \hat{M} R^T =0.
\label{eq:exact-seesaw}
\end{equation}
By taking the $ee$ elements of the terms  in the  left-hand side of  Eq.(\ref{eq:exact-seesaw}), one can express the 
masses of the sterile states in terms of the active neutrino masses and other mixing parameters. We take up this calculation in the next section.
\subsection{Parameterization of the \texorpdfstring{$3+3$}{} model}
\label{sec:Parametrisation}
 In this framework, the neutral fermion spectrum comprises six mass eigenstates ($\nu_1, \nu_2, \nu_3, N_1, N_2 ,N_3$). Thus, the leptonic mixings are encoded within a $6\times6$ unitary mixing matrix $\mathcal{U}$. Following \cite{Xing:2011ur, Xing:2007zj},for instance, the matrix $\mathcal{U}$ can be formed in terms of 15 rotational angles ($\theta_{ij}$) and 15 phases ($\phi_{ij}$).  Expressing $\mathcal{U}$  as a product of $15$ 
 two-dimensional rotational matrices in a six dimensional complex space we can write
\begin{equation}
{\mathcal{U}} = O_{56}O_{46}O_{45}O_{36}O_{26}O_{16} O_{35} O_{25}O_{15}O_{34}O_{24}O_{14}O_{23}O_{13}O_{12}.
\label{eq:U-expression}
\end{equation}
The above rotations are of the following form (exemplified by $O_{56}$):
\begin{equation}
O_{56}=
\begin{pmatrix}
1&0&0&0&0&0\\0&1&0&0&0&0 
\\0&0&1&0&0&0\\0&0&0&1&0&0\\0&0&0&0&c_{56}&s_{56}e^{-i\phi_{56}}
\\0&0&0&0&-s_{56}e^{i\phi_{56}}&c_{56}\end{pmatrix},
\label{eq:rot-matrix}
\end{equation}
where $c_{ij} = \cos \theta_{ij}$  and $s_{ij} = \sin \theta_{ij}$,  $\theta_{ij}$ are the  mixing angles and  $\phi_{ij}$ represent the CPV phases, respectively. Here, we have taken $\phi_{ij} = \delta_{ij} + \rho_{ij}$; $\delta_{ij}$ and $\rho_{ij}$ as the Dirac-type and Majorana-type phases with ($\delta_{ij},\rho_{ij} \in [0,2\pi]$). Further, the non-unitarity effect, being an inherent feature due to the admixture of active-sterile states, can be visualized from each $3\times3$ block of $\mathcal{U}$.

Now, accounting for the $ee$ elements from the exact seesaw relation shown in Eq.(\ref{eq:exact-seesaw}), we come up with the following relation:
\begin{equation}
\sum_{i=1}^3 m_i V^2_{ei} + \sum_{k=1}^3 M_k R^2_{ek} = 0.
\label{eq:ee-element}
\end{equation}
Using Eqs. (\ref{eq:exact-seesaw}) and (\ref{eq:ee-element}) three equations, linear in  $M_1, M_2$, and $M_3$, can be written as
\begin{equation}
\sum_{k=1}^3 M_k {\rm Re} R^2_{ek} =  -\sum_{i=1}^3 m_i {\rm Re} V^2_{ei},
\label{eq:linear-1}
\end{equation}
\begin{equation}
\sum_{k=1}^3 M_k {\rm Im} R^2_{ek} = -\sum_{i=1}^3 m_i {\rm Im}V^2_{ei},
\label{eq:linear-2}
\end{equation}
\begin{eqnarray}
\left(\sum_i R_{i1}^2\right)M_1 + \left(\sum_i R_{i2}^2\right)M_2 + \left(\sum_i R_{i3}^2\right)M_3 &=&\nonumber \\
-\left(\left(\sum_i V_{i1}^2\right)m_1 + \left(\sum_i V_{i2}^2\right)m_2 + \left(\sum_i V_{i3}^2\right)m_3\right).
\label{eq:linear-3}
\end{eqnarray}
Equations (\ref{eq:linear-1}) and (\ref{eq:linear-2}) are obtained by equating the real and
imaginary parts of left-hand side of Eq.(\ref{eq:ee-element}) to zero. Equation (\ref{eq:linear-3})
is obtained by relating the traces of the two terms in the left-hand side of Eq.(\ref{eq:exact-seesaw}).  
Performing a simple calculation, we can obtain the analytical solutions for the masses
of the sterile states as follows:

 \begin{equation}
    M_1=\frac{d_1(b_3 c_2 - b_2 c_3) + d_2(b_1 c_3 - b_3 c_1) + d_3(b_2 c_1 - b_1 c_2)}{a_1(b_2c_3 - b_3c_2) - a_2(b_1c_3 - b_3c_1) + a_3(b_1c_2 - b_2c_1)},
    \label{eq:sol-mass1}
\end{equation}
 \begin{equation}
    M_2=\frac{d_1(a_2 c_3 - a_3 c_2) + d_2(a_3 c_1 - a_1 c_3) + d_3(a_1 c_2 - a_2 c_1)}{a_1(b_2c_3 - b_3c_2) - a_2(b_1c_3 - b_3c_1) + a_3(b_1c_2 - b_2c_1)},
    \label{eq:sol-mass2}
\end{equation}
 \begin{equation}
   M_3=\frac{d_1(a_3 b_2 - a_2 b_3) + d_2(a_1 b_3 - a_3 b_1) + d_3(a_2 b_1 - a_1 b_2)}{a_1(b_2c_3 - b_3c_2) - a_2(b_1c_3 - b_3c_1) + a_3(b_1c_2 - b_2c_1)},
   \label{eq:sol-mass3}
\end{equation}
\vspace{0.5pt}
where 
 \begin{equation*}
    \begin{split}
         a_1={\rm Re} R_{e1}^2,\quad b_1={\rm Re} R_{e2}^2,\quad c_1= {\rm Re}R_{e3}^2,\quad d_1=\sum_{i=1}^3 m_i {\rm Re} V^2_{ei},\\
          a_2={\rm Im} R_{e1}^2,\quad b_2={\rm Im} R_{e2}^2,\quad c_2={\rm Im} R_{e3}^2,\quad d_2=\sum_{i=1}^3 m_i {\rm Im} V^2_{ei},\\
         a_3=\sum_{i=1}^3 R_{i1}^2,\;\quad b_3=\sum_{i=1}^3 R_{i2}^2,\quad c_3=\sum_{i=1}^3 R_{i3}^2,\quad d_3=\sum_{i,j=1}^3 V_{ij}^2 m_j.
    \end{split}
    \label{eq:abbrieviation}
\end{equation*}
It can be observed that, using the parametric expressions for V and R from Eq.(\ref{eq:rot-matrix}), the analytical expressions for $M_1, M_2$, and $M_3$ now become a function of the mixing angles, CPV phases, and active neutrino masses. This in turn provides a platform where the impact of these observables can be explored on the effective mass of the $0\nu \beta \beta$ decay process.
\section{Neutrinoless double beta decay }
\label{sec:ovbb}
 Neutrinoless double beta decay is a second - order weak process which corresponds to the transition from a nucleus $(A, Z)$ to its isobar $(A, Z+2)$ with the emission of two electrons, $(A,Z) \rightarrow (A,Z+2) + 2e^-$. If observed, it will confirm the Majorana nature of neutrinos and lepton number violation.  In the SM framework, the $0\nu\beta\beta$ decay is forbidden, while it is allowed in BSM scenarios.
 \begin{figure}[htb]
    \centering
    \includegraphics[width=0.5\linewidth,height=4.95cm]{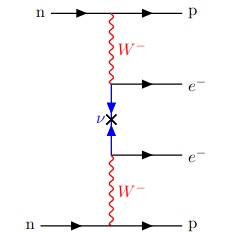} 
    \includegraphics[width=0.48\linewidth,height=5cm]{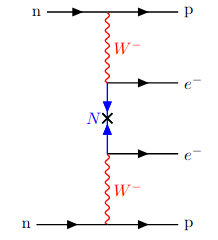} 
    \caption{Feynman diagrams representing $0\nu\beta\beta$ decay mediated by the active neutrinos (left) and RH sterile neutrinos (right).}
    \label{fig:feyn-diagram}
\end{figure}
\subsection{Standard three - neutrino framework}
\label{sm-framework}
 Assuming the decay process to be mediated by light Majorana neutrinos, the inverse half-life period $[t^{0\nu\beta\beta}_{1/2}]^{-1}$ of the decay process can be calculated with \cite{Brofferio:2019yoc, Rodejohann:2011mu, DellOro:2016tmg,  Belley:2023btr}
 \begin{equation}
        [t^{0\nu\beta\beta}_{1/2}]^{-1} = 
        G^{0\nu\beta\beta}\left|M^{0\nu\beta\beta}\right|^2 \frac{\lvert m_{ee} \lvert^2}{m_e^2} ,
        \label{eq:timeperiod}
 \end{equation}
 where  $G^{0\nu\beta\beta}$ is the phase space factor, $m_e$ is the mass of electron,   $M^{0\nu\beta\beta}$ is the nuclear matrix element, and $\lvert m_{ee} \lvert$ is the effective neutrino mass that indicate the absolute neutrino mass scale and mass hierarchy.
 In the context of three generation of neutrinos, the effective mass is given by (see e.g.,\cite{Elliott:2002xe, Penedo:2018kpc, Dolinski:2019nrj})
 \begin{equation}
     \lvert m_{ee} \lvert = \left |\sum_{j=1}^3 m_j U^2_{ej}\right |.
     \label{eq:eff-mass}
    \end{equation}
Here, the sum runs over masses of the three active neutrinos, $m_j$, and the elements of the active neutrino mixing matrix $U_{ej}$, where $U$ is a Pontecorvo–Maki–Nakagawa–Sakata (PMNS) matrix. In the standard parameterization, it is given as  
\begin{equation}
           U = \begin{pmatrix} \substack{ c_{12}c_{13}} &\substack{ s_{12}c_{13}} &\substack{ s_{13}e^{-i\delta}} \\
           \substack{-s_{12}c_{23} - c_{12}s_{13}s_{23}e^{i\delta}} &\substack{ c_{12}c_{23}-s_{12}s_{13}s_{23}e^{i\delta}} &\substack{ c_{13}s_{23}} \\
           \substack{s_{12}s_{23}-c_{12}s_{13}c_{23}e^{i\delta}} &\substack{ -c_{12}s_{23}-s_{12}s_{13}c_{23}e^{i\delta}} & \substack{c_{13}c_{23}} 
           \end{pmatrix}P.
           \label{eq:standard-U}
        \end{equation}
Here, the symbols $c_{ab}$ and $s_{ab}$ stand for $\cos{\theta_{ab}}$ and $\sin{\theta_{ab}}$, respectively with $\theta_{ab}$ being the mixing angles, and $P=\\
\textrm{diag}(1,e^{i\rho},e^{i(\beta+\delta)})$ is a diagonal matrix containing the Majorana-type CPV phases $ \rho$ and $\beta$ and the Dirac-type CPV phase  $\delta$. Both the Dirac and Majorana CPV phases have the variation limit in the interval [0, $2\pi$].
Now, according to the standard parameterization given in Eq.(\ref{eq:standard-U}), the effective neutrino mass can be written as
 \begin{equation}
   \lvert m_{ee} \lvert= \lvert m_1c_{12}^2c_{13}^2 + m_2s_{12}^2c_{13}^2e^{2i\rho} + m_3s_{13}^2e^{2i\beta}\lvert.
   \label{eq:eff-mass-exp}
\end{equation}
Further, the effective mass can have different values depending on the mass ordering followed by the neutrino mass states, such as 
\begin{itemize}
\item The normal hierarchy (NH) corresponds to the arrangement $m_3>m_2>m_1$ with
\begin{equation}
 \begin{split}
    m_2 = \sqrt{m_1^2 + \Delta m_{21}^2}, \qquad
    m_3 = \sqrt{m_1^2 + \Delta m_{31}^2}.
\end{split}  
\label{eq:nh}
\end{equation}
\item The invented hierarchy (IH) corresponds to the arrangement $m_2>m_1>m_3$ with 
\begin{equation}
 \begin{split}
    m_1 = \sqrt{m_3^2 -
    \Delta m_{32}^2-\Delta m_{21}^2}, \qquad
    m_2 = \sqrt{m_3^2 - \Delta m_{32}^2}.
\end{split}  
\label{eq:ih}
\end{equation}
\end{itemize}
\begin{figure}[htb]
 \centering
 \includegraphics[width=0.45\textwidth]{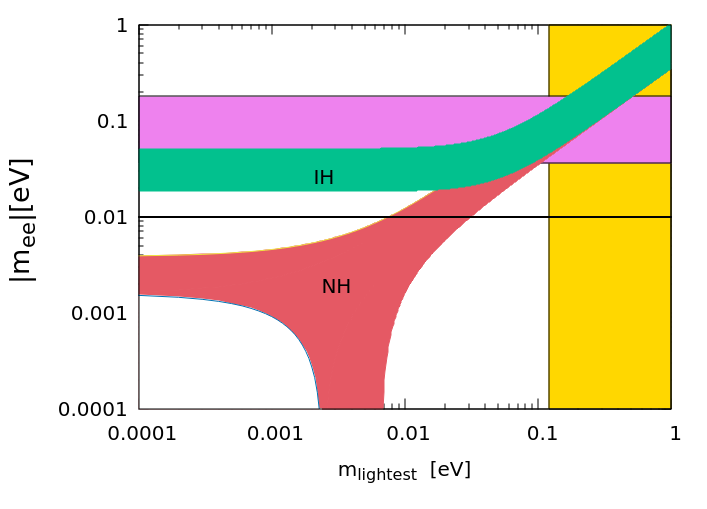}
 \caption{Effective mass as a function of the lightest neutrino mass within the standard active neutrino framework, both for the NH and IH, where the oscillation parameters are taken in $1\sigma$ range. The yellow region depicts the cosmological disfavored region. The pink region represents the experimental limits from  KamLAND-Zen($ 0.036-0.156$ eV) \cite{KamLAND-Zen:2022tow} + GERDA($0.079-0.18$ eV) \cite{GERDA:2020xhi}. 
The black line signifies the projected sensitivity of $\lvert m_{ee}  \lvert$ in future experiments such as CUPID, nEXO, and LEGEND.}
 \label{fig:mee-sm}
\end{figure}
Figure(\ref{fig:mee-sm}) represents the parameter space of $\lvert m_{ee}  \lvert$ as a function of the lightest neutrino mass in a standard three-active-neutrinos picture for the NH and IH mass schemes.
The results of our analysis in the subsequent section involving
sterile neutrinos can be interpreted by comparing the new contributions  with the standard-three-active neutrinos picture. Using data from neutrino oscillation and cosmology, one can 
obtain predictions for possible values of $\lvert m_{ee}  \lvert$\cite{Barabash:2018fun}. In the standard three - light - neutrino picture, where the $0\nu\beta\beta$ decay process is mediated by active neutrinos,
for the NH, $\lvert m_{ee}  \lvert \approx (0 - 30)\times 10^{-3}$ eV depending upon different values of $m_{\rm lightest}$.
 For $m_{\rm lightest} \leq 10^{-3}$ eV, $\lvert m_{ee}  \lvert \approx (1 - 4)\times 10^{-3}$ eV. For $m_{\rm lightest}$ in the range $(1 - 10)\times 10^{-3}$ eV, there is a possibility of strong cancellation in the  $\lvert m_{ee}  \lvert$ due to the contributions of Majorana phases. For the IH, the effective mass $\lvert m_{ee}  \lvert \approx (10 - 50)\times 10^{-3}$ eV for all allowed values of  $m_{\rm lightest}$.
If $0\nu \beta \beta$ decay is detected in the future, it would be possible to make definitive predictions on the value of
$\lvert m_{ee}  \lvert$. 
Global analysis of neutrino oscillation data prefers
that the NH is preferred over the IH \cite{Capozzi:2017ipn, Esteban:2020cvm}  at $3.5 \sigma$ level \cite{DeSalas:2018rby}. Assuming neutrinos are Majorana particles, a limit on 
$\lvert m_{ee}  \lvert$ below $14$ MeV would rule out the IH scheme.
The future ton-scale experiments like CUPID \cite{CUPID:2019imh,Armengaud:2019loe}, LEGEND \cite{LEGEND:2017cdu}, and nEXO \cite{nEXO:2017nam,nEXO:2018ylp} are going to probe  $\lvert m_{ee}  \lvert \le 0.01$ eV.
If $\lvert m_{ee}\lvert$ is not registered in the above stated ranges, the presence of new physics contributions other than the light neutrinos can
register very low values of $\lvert m_{ee}  \lvert$.
\subsection{Neutrinoless double beta decay in the presence of  sterile neutrinos in light of the exact seesaw relation}
\label{sec:ovbb-sterile}
In this section, we study the effect of the combined contribution of active and sterile neutrinos
on $\lvert m_{ee}  \lvert$. While studying the contributions from sterile states, we
make use of the Eqs. (\ref{eq:sol-mass1}- \ref{eq:sol-mass3}), which arise as a consequence of an exact seesaw relation.
The effect of the addition of $N$ number of sterile neutrinos on $\lvert m_{ee} \rvert$ has been reported in Ref.\cite{Blennow:2010th}, as a special emphasis on the dependence of mass of 
sterile states on the $0\nu\beta\beta$ decay rate. In Ref.\cite{Abada:2014nwa}, the same idea  has  been incorporated in an inverse seesaw model. In Ref.\cite{Blennow:2010th}, the authors also show that the contribution of active neutrinos should always be taken into account when deriving bounds on additional parameters of the model. Accordingly, inclusion of three sterile states  leads to modification of the effective neutrino mass formula given in Eq.(\ref{eq:eff-mass}). So, taking into account both active and sterile neutrinos mediating the  $0\nu\beta\beta$ decay process and using the parameterization given in Eq.(\ref{eq:curl-U}), the effective neutrino mass can be expressed as \cite{Abada:2021zcm, Abada:2014nwa, Abada:2018qok}:
\begin{equation}
\lvert m_{ee}  \lvert = \left |\sum_{i=1}^3 m_i V^2_{ei} + \sum_{k=1}^3 R^2_{ek}\frac{M_k}{p^2 - M_k^2}p^2\right |,
\label{eq:eff-mass-sterile}
\end{equation}
where $p^2=-(100 ~{\rm MeV})^2$ is the virtual momentum exchanged in the process. 
As the only neutrino associated with the decay is the electron neutrino $\nu_e$, only the  $ei$-elements of the neutrino mixing matrix $\mathcal{U}$ are present in the above 
expression.
\subsection{Analysis of parameter space}
\label{sec:parameter}
 we consider in our analysis the three sterile states to be non-degenerate so that their mixing with active ones become relatable and the effect of CPV phases becomes prominent.  So far, $\lvert m_{ee}  \lvert$ is a function of $15$ mixing angles, $15$ phases, and the masses of active neutrinos. These $15$ mixing angles include $3$ active-active, $3$ sterile-sterile, and $9$ active-sterile mixing angles. Out of these, we consider the sterile-sterile mixing angles to be zero; $\theta_{45}=\theta_{46}=\theta_{56}=0$. The values of the $3$ active-active mixing angles ($\theta_{12},\theta_{13}$, and $\theta_{23}$) are known from oscillation experiments \cite{Esteban:2020cvm,deSalas:2020pgw,Capozzi:2021fjo,Gonzalez-Garcia:2021dve}. Thus, we are left with only nine active-sterile mixing angles, $\theta_{14}, \theta_{24}, \theta_{34}, \theta_{15}$, 
$\theta_{25},\theta_{35}, \theta_{16}, \theta_{26}, \textrm{and} \theta_{36}$, whose contribution has to be examined.

Similarly, out of the $15$ phases, we mark those phases as unphysical which are connected to the sterile-sterile mixing, i.e., $\phi_{45}, \phi_{46},\textrm{and} \phi_{56}$. From the remaining $12$ phases, using the expressions of $M_1, M_2$ and $M_3$ given in Eqs. (\ref{eq:sol-mass1} -\ref{eq:sol-mass3}), we have selected five phases which are found to be more important for their direct contribution to those masses. They are $\phi_{12},\phi_{13}, \phi_{14}, \phi_{15}$, and $\phi_{16}$. The masses, $M_1, M_2$, and $M_3$ also depend on the mass of the active neutrinos. So, adhering to a particular hierarchical pattern, as shown in Eqs. (\ref{eq:nh}) and (\ref{eq:ih}), the mass of the  sterile states can be expressed as a function of the  lightest neutrino mass. In this way, the parameter space for the analysis
is left with nine mixing angles and five CPV phases along with the lightest neutrino mass.
\section{Results and discussions}\label{sec:result}
We discuss in this section the possible outcomes that are obtained through the selected parameter space.
In the previous subsection, we selected five phases $\phi_{1i}$; $i=2,3,4,5,6$ (which are actually the combination of Dirac and Majorana phases) to be of primary interest for our study. Though they all have an independent variation from 0 to $2\pi$, in order to have finite masses, the meticulous observations from the analytical expressions of $M_1, M_2$ and $M_3$ demand the following constraint:
\begin{equation}
\phi_{14} \ne \phi_{15} \ne \phi_{16} \ne \frac{n\pi}{2},\; n=0,1,2,3, \cdots.
\label{eq:phase-constraint}
\end{equation}
 One of the experimental verifications of the type-I seesaw mechanism demands that
the RH neutrinos should be discovered at the  
direct searches such as discussed in Refs. \cite{Bolton:2019pcu, Abdullahi:2022jlv}. 
The sensitivity of the searches requires their masses must be $\mathcal{O}(1)$ TeV 
or smaller and their coupling to the charged leptons must not be too small
(for example, Figs (26), (27), and (28) of Ref. \cite{Abdullahi:2022jlv} and 
Figs (6), (9), (10), (11), and (12) of Ref. \cite{Bolton:2019pcu}). Keeping these
criteria in mind, we explore the parameter space of taking three sets of benchmark
values of the active-sterile mixing angles such as (see for example \cite{Abada:2021zcm}) the following: 
\begin{itemize}
 \item Set 1: $\theta_{1j}\approx10^{-5}$, $\theta_{2j}\approx10^{-4}$, $\theta_{3j}\approx10^{-3}$,
 \item Set 2:  $\theta_{1j}\approx10^{-4}$, $\theta_{2j}\approx10^{-3}$, $\theta_{3j}\approx10^{-2}$, 
 \item Set 3: $\theta_{1j}\approx10^{-3}$, $\theta_{2j}\approx10^{-2}$, $\theta_{3j}\approx10^{-1}$.
\end{itemize} 
As $0\nu\beta\beta$ decay is sensitive to the mass of sterile neutrinos and active-sterile mixing angles,
we examine the impact of the parameter space on effective neutrino mass as a function of $m_{\rm lightest}$
and $\sum m$. ( A correlation between $0\nu\beta\beta$ other other direct and indirect constraints from
lepton number - conserving and number - violating processes with one generation of an active neutrino
and two sterile neutrinos was studied in Ref. \cite{Bolton:2019pcu}).
Further, the neutrino oscillation data and precision electroweak data constrain the unitarity violation of $V$ above the order of $\mathcal{O}(10^{-2})$ \cite{Xing:2007zj, Xing:2009in}. Hence, this small non-unitarity effect of $V$ in turn implies the active-sterile mixing to be small.
A discussion on the compatibility of the selected parameter space with the constraints from
non-unitarity is provided in section (\ref{sec:nonunitary-range}).
\begin{figure*}[htbp]  
    \centering
    \begin{subfigure}{0.3\textwidth}
        \centering
        \includegraphics[width=\linewidth]{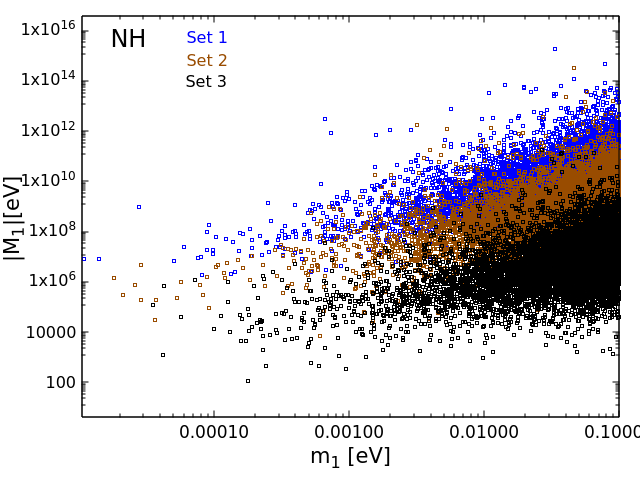}
        \caption{}
        \label{fig:sub1}
    \end{subfigure}
    \hfill
    \begin{subfigure}{0.3\textwidth}
        \centering
        \includegraphics[width=\linewidth]{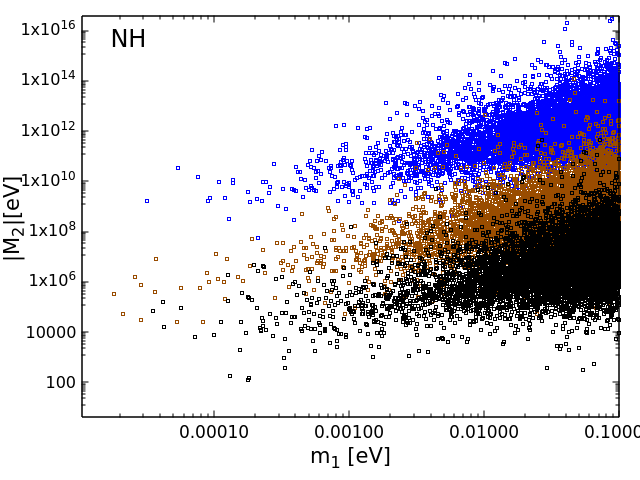}
        \caption{}
        \label{fig:sub2}
    \end{subfigure}
    \hfill
    \begin{subfigure}{0.3\textwidth}
        \centering
        \includegraphics[width=\linewidth]{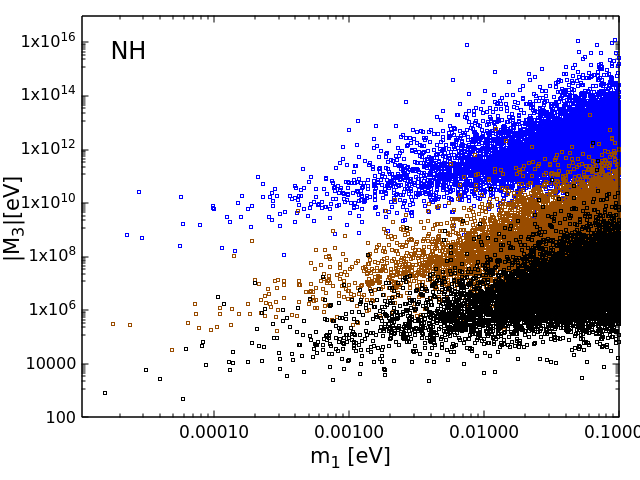}
        \caption{}
        \label{fig:sub3}
    \end{subfigure}
    \medskip
    \begin{subfigure}{0.3\textwidth}
        \centering
        \includegraphics[width=\linewidth]{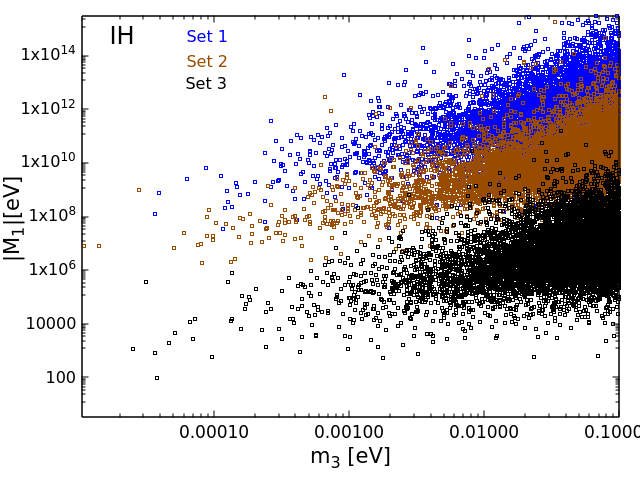}
        \caption{}
        \label{fig:sub4}
    \end{subfigure}
    \hfill
    \begin{subfigure}{0.3\textwidth}
        \centering
        \includegraphics[width=\linewidth]{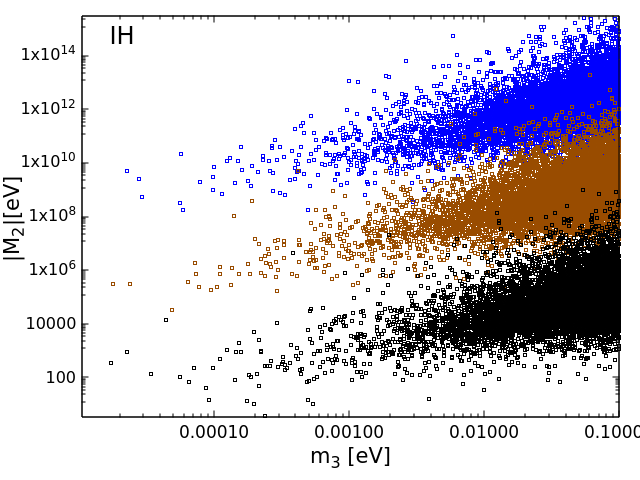}
        \caption{}
        \label{fig:sub5}
    \end{subfigure}
    \hfill
    \begin{subfigure}{0.3\textwidth}
        \centering
        \includegraphics[width=\linewidth]{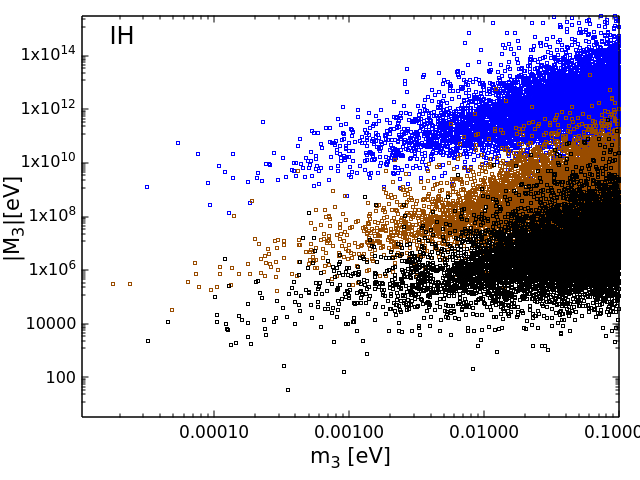}
        \caption{}
        \label{fig:sub6}
    \end{subfigure}
    \caption{ Numerically obtained ranges for masses of sterile states $M_1, M_2$, and $M_3$ against the mass of the lightest active neutrino state using Eqs. (\ref{eq:sol-mass1} - \ref{eq:sol-mass3}). The blue, brown, and black 
    points correspond to the Sets 1, 2, and 3 of 
    combination of active-sterile mixing angles, respectively. The upper (lower) panel
 is for the NH (IH).}
    \label{fig:sterile-mass}
\end{figure*}

\begin{figure*}
    \centering
    \begin{subfigure}{0.3\textwidth}
        \centering
        \includegraphics[width=\linewidth]{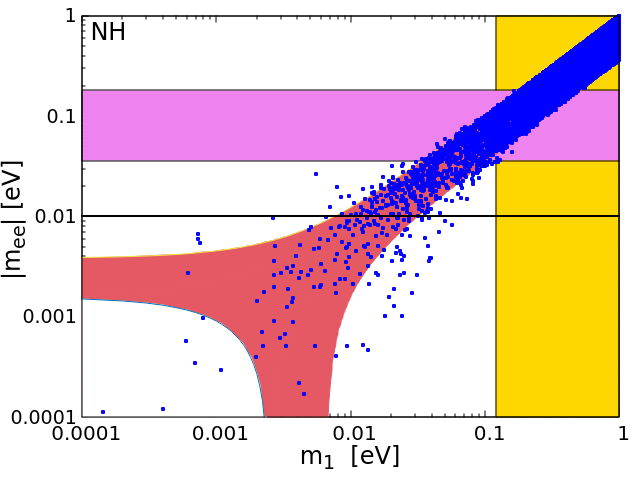}
        \caption{}
        \label{fig:sub7}
    \end{subfigure}
    \hfill
    \begin{subfigure}{0.3\textwidth}
        \centering
        \includegraphics[width=\linewidth]{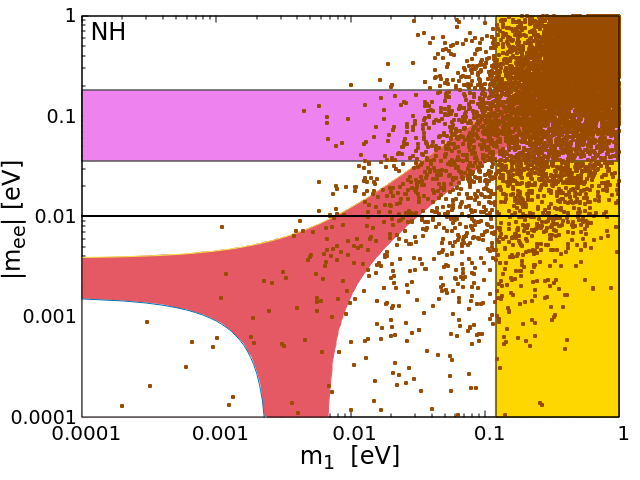}
        \caption{}
        \label{fig:sub8}
    \end{subfigure}
    \hfill
    \begin{subfigure}{0.3\textwidth}
        \centering
        \includegraphics[width=\linewidth]{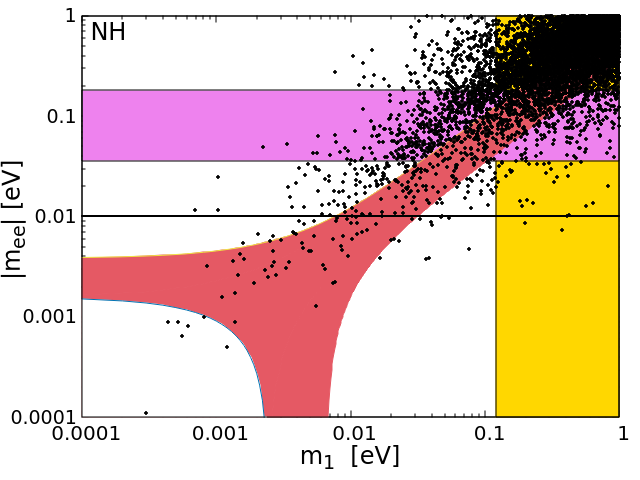}
        \caption{}
        \label{fig:sub9}
    \end{subfigure}
    \medskip
    \begin{subfigure}{0.3\textwidth}
        \centering
        \includegraphics[width=\linewidth]{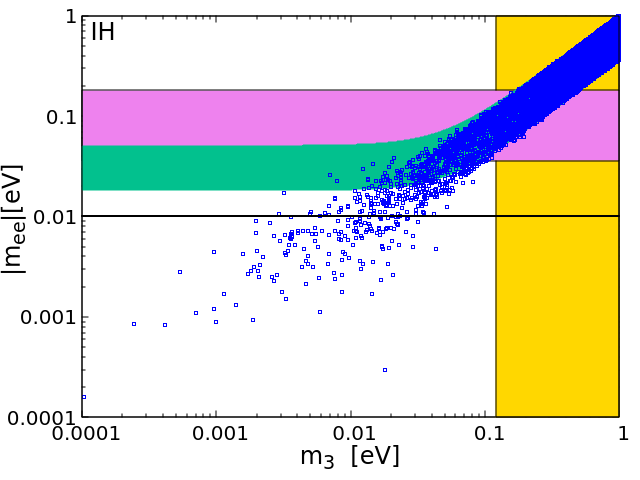}
        \caption{}
        \label{fig:sub10}
    \end{subfigure}
    \hfill
    \begin{subfigure}{0.3\textwidth}
        \centering
        \includegraphics[width=\linewidth]{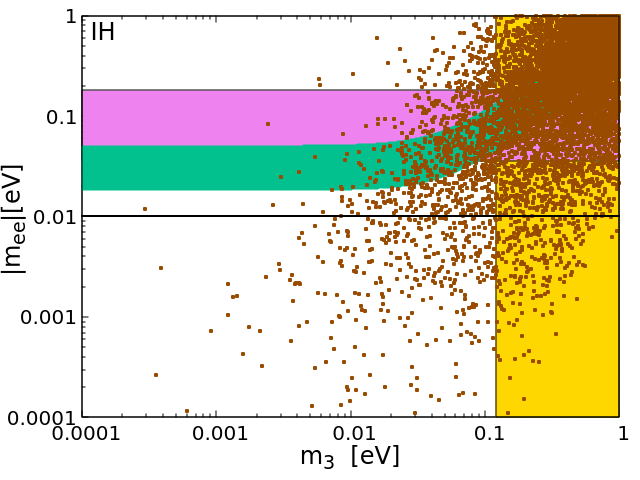}
        \caption{}
        \label{fig:sub11}
    \end{subfigure}
    \hfill
    \begin{subfigure}{0.3\textwidth}
        \centering
        \includegraphics[width=\linewidth]{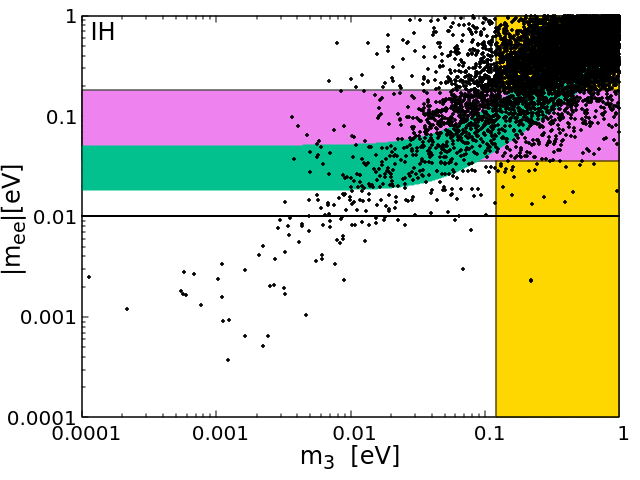}
        \caption{}
        \label{fig:sub12}
    \end{subfigure}
    \caption{ Effective mass against lightest neutrino mass. The colored dots show the effective neutrino mass numerically obtained from the combined contribution of active and sterile neutrinos.  The red (green) regions represent the standard active neutrino contribution to the effective mass for the NH (IH).
The yellow regions in all the figures correspond to the cosmological disfavored region. The pink region represents the experimental limits from  KamLAND-Zen ($ 0.036-0.156$ eV)  \cite{KamLAND-Zen:2022tow} + GERDA ($0.079-0.18$ eV)  \cite{GERDA:2020xhi}. 
The black line signifies the projected sensitivity of $\lvert m_{ee}  \lvert$ in future experiments like CUPID, nEXO, and LEGEND. The upper (lower) panel is for the NH (IH) of active neutrinos. The left, middle, and right columns correspond to the combination 
of active-sterile mixing angles of Sets 1, 2, and 3, respectively.}
    \label{fig:mee-bsm}
\end{figure*}

\begin{figure*}
    \centering
     \begin{subfigure}{0.3\textwidth}
        \centering
        \includegraphics[width=\linewidth]{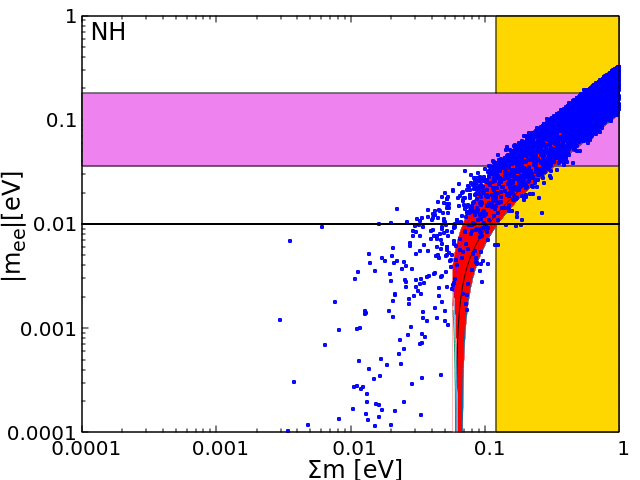}
        \caption{}
        \label{fig:sub13}
    \end{subfigure}
    \hfill
    \begin{subfigure}{0.3\textwidth}
        \centering
        \includegraphics[width=\linewidth]{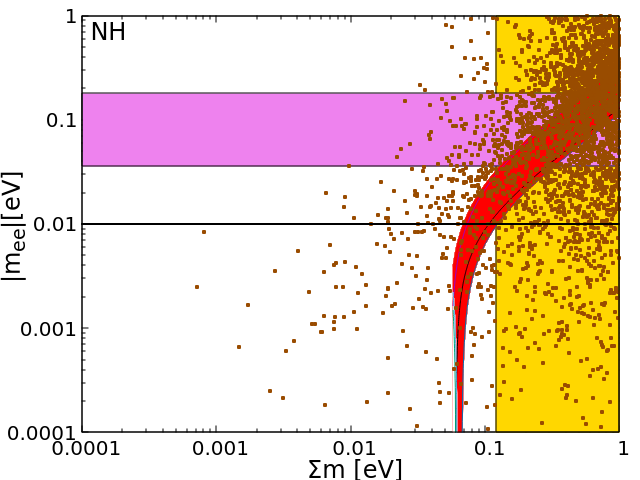}
        \caption{}
        \label{fig:sub14}
    \end{subfigure}
    \hfill
    \begin{subfigure}{0.3\textwidth}
        \centering
        \includegraphics[width=\linewidth]{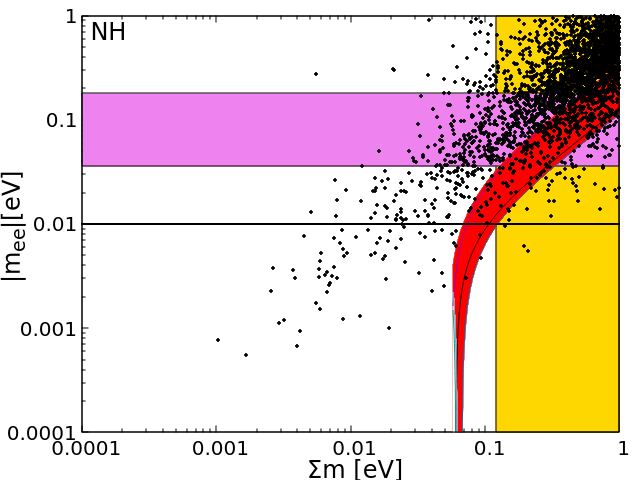}
        \caption{}
        \label{fig:sub15}
    \end{subfigure}

    \medskip

    \begin{subfigure}{0.3\textwidth}
        \centering
        \includegraphics[width=\linewidth]{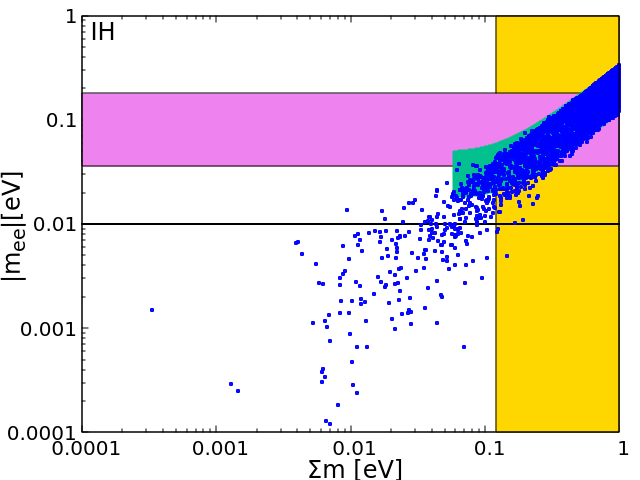}
        \caption{}
        \label{fig:sub16}
    \end{subfigure}
    \hfill
    \begin{subfigure}{0.3\textwidth}
        \centering
        \includegraphics[width=\linewidth]{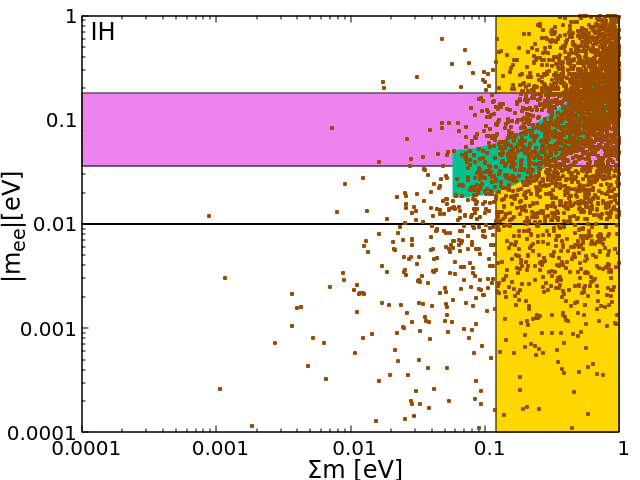}
        \caption{}
        \label{fig:sub17}
    \end{subfigure}
    \hfill
    \begin{subfigure}{0.3\textwidth}
        \centering
        \includegraphics[width=\linewidth]{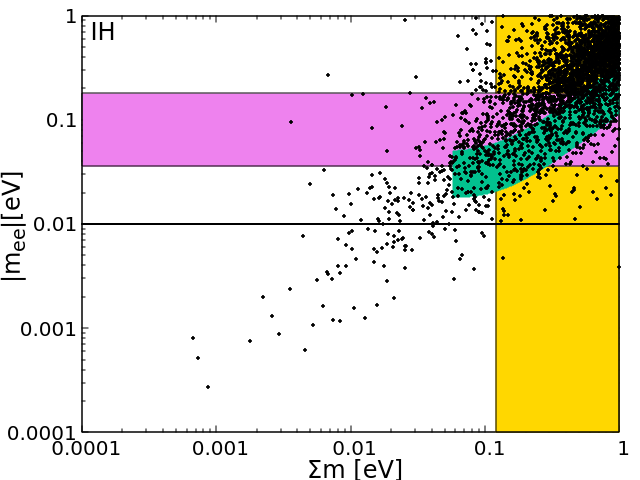}
        \caption{}
        \label{fig:sub18}
    \end{subfigure}
    \caption{ Effective mass against the sum of light neutrino masses:
 the red (green) regions represent the standard active neutrino contribution to the effective mass for the 
NH (IH). The yellow regions in all the figures correspond to the cosmological disfavored region.  The pink region presents the experimental limits from KamLAND-Zen ($0.036 - 0.156$ eV ) \cite{KamLAND-Zen:2022tow} $+$ GERDA ($0.079-0.18$ eV) \cite{GERDA:2020xhi}. The black line signifies the projected sensitivity of $\lvert m_{ee}  \lvert$ in future experiments like CUPID, nEXO, and LEGEND. The upper (lower) panel is for the NH (IH) of active neutrinos. The left, middle, and right columns correspond to the combination 
of active-sterile mixing angles of Sets 1, 2, and 3, respectively.}
    \label{fig:mee-sum}
\end{figure*}

Accomplished with the above conditions for phases and mixing angles, we numerically
obtain the effective mass of electron neutrino $\lvert m_{ee} \rvert$ using Eq.(\ref{eq:eff-mass-sterile}). The present and future experimental searches for $0\nu\beta\beta$ decay along with cosmological observations act as the 
guiding light in testing BSM scenarios in this context.
We analyze the values, thus obtained, by making a comparison  between the combined contribution and active neutrino contribution to $\lvert m_{ee} \rvert$ against the lightest neutrino mass and sum of light neutrino masses in Figs. (\ref{fig:mee-bsm}) and (\ref{fig:mee-sum}), respectively. 
While calculating $\lvert m_{ee} \rvert$ using Eq.(\ref{eq:eff-mass-sterile}), numerically, first we have calculated the masses $M_1, M_2$, and $M_3$ of the sterile states 
using Eqs. (\ref{eq:sol-mass1} - \ref{eq:sol-mass3}). They are plotted in Fig.(\ref{fig:sterile-mass}) as a function of the mass of the lightest active neutrino and mixing angles and phases of the $(6\times 6)$ mixing matrix $\mathcal{U}$. The masses of the involved sterile states are found in the $\sim$ keV - TeV range. 
 In Fig.(\ref{fig:sterile-mass}), the upper panel corresponds to the NH and the lower panel corresponds to the IH cases for active neutrinos. In each panel, the impact of the active-sterile mixing angles 
on $M_i$'s can be observed for the three benchmark sets. It can be observed from Fig.(\ref{fig:sterile-mass}) that for Set 1, the 
masses can reach up to orders of TeV, and for Set 3, they can reach up to orders of GeV.
Depending upon their masses, the sterile states along with active states can show some distinct features in their contribution to the amplitude of 
$0\nu\beta\beta$ decay, which was shown in Ref.\cite{Blennow:2010th}. 
The masses of the sterile states can be all light or all heavy or a mixture of
both heavy and light. For different cases, the combined contribution of active and sterile states to the the amplitude of $0\nu\beta\beta$ decay is summarized in Appendix \ref{app:nme}. 

In Fig.(\ref{fig:mee-bsm}), the values for  $\lvert m_{ee}\lvert$, from the combined contribution of active and sterile states, are shown for the NH and IH.
 The colored dots represent the numerically obtained values of 
$\lvert m_{ee}\lvert$. 
The left, middle, and right columns corresponds to the combination 
of active-sterile mixing angles of Sets 1, 2. and 3, respectively, 
with the color code corresponding to Sets 1, 2, and 3 similar to 
Fig.(\ref{fig:sterile-mass}).  As a general observation, one can observe that the parameter space of $\lvert m_{ee}\lvert$ versus $m_{\rm lightest}$ 
is sensitive to the combination of active-sterile mixing angles.
From Fig.(\ref{fig:mee-bsm}), the lower bound on $\lvert m_{ee}\lvert$ corresponding to the lightest mass for both the NH and IH can be revealed. The lower limit on the lightest active neutrino mass in the study of $\lvert m_{ee}\lvert$  is of the order $\mathcal{O}(10^{-4})$, as has been derived in the context of $SO(10)$ in
Ref.\cite{Buccella:2017jkx} and in left-right symmetric model presented in Ref.\cite{BhupalDev:2013ntw}.

The value of the lightest neutrino mass, for which the combined contribution of active and sterile neutrinos saturate the experimental bound, is lower than that of standard light neutrino contribution for the NH. 
For the cancellation region corresponding to the NH, there is some parameter space available where the combined contribution shows enhancement in $\lvert m_{ee} \rvert$. This could be probed by future ton-scale experiments, CUPID \cite{CUPID:2019imh,Armengaud:2019loe}, LEGEND \cite{LEGEND:2017cdu}, and nEXO \cite{nEXO:2017nam,nEXO:2018ylp}  which aim for sensitivity of $\lvert m_{ee} \rvert$ as low as $0.01$ eV. For the NH,
for lower values of $m_1$ ($\mathcal{O} (10^{-4})
$ eV), the standard contribution dominates over the combined contribution. It shows the mutual cancellation between the contribution from  light and sterile states to $\lvert m_{ee} \rvert$. The same behavior
is also seen in case of the IH, where for lower values of  $m_3$ ($\mathcal{O} (10^{-3})$ eV), the standard contribution is dominant over the combined contribution.
However, unlike the
standard active neutrino contribution, the combined contribution of active and sterile states to $\lvert m_{ee}\lvert$ increases with the lightest mass for both the NH and IH.

we show in Fig.(\ref{fig:mee-sum}) the combined contribution to $\lvert m_{ee} \rvert$  against the sum of light neutrino masses for the NH and IH. The colored dots represent the numerically
obtained values of $\lvert m_{ee}\rvert$ coming from combined contribution of active and sterile neutrinos to the $0\nu\beta\beta$ decay process. 
Some general observations can be made from Fig.(\ref{fig:mee-sum}); the combined contribution to the $0\nu\beta\beta$ decay process offers more parameter space for the lower values
of $\sum m_i$.  Also, the combined contribution saturates the present and future laboratory bound
for lower values of $\sum m_i$ except for Set 1 for IH.
The future ton-scale experiments like CUPID \cite{CUPID:2019imh,Armengaud:2019loe}, LEGEND \cite{LEGEND:2017cdu}, and nEXO \cite{nEXO:2017nam,nEXO:2018ylp} are going to probe $\lvert m_{ee} \rvert \le 0.01$ eV. The future cosmological observations are going to probe the sum of the active neutrino masses in the range $0.03 - 0.06$ eV \cite{Amendola:2016saw,SPT-3G:2019sok,CMB-S4:2022ght}.
With the projected sensitivity of the sum of active neutrino masses and if 
$0\nu\beta \beta$ decay is registered with $\lvert m_{ee}\rvert \lesssim 0.01$ eV,
the standard three - neutrino contribution to $\lvert m_{ee}\rvert$ will be in conflict with the observations. But the combined contribution will still offer some parameter
space to study new physics contributions to the $0\nu\beta\beta$ decay process.
Global fit \cite{Capozzi:2017ipn,Esteban:2020cvm} of neutrino oscillation data shows the NH is favored over the IH at the level of $\sim 2\sigma$.
Contemplating a situation where a global fit of neutrino oscillation data
disfavors the IH, and future sensitivity of $0\nu \beta \beta$ reaches $\lvert m_{ee}\rvert \sim 0.01$ eV, sterile neutrinos will leave some amount of parameter space
for future search. Also the data from cosmology could possibly impose a lower limit
on the lightest neutrino mass.

The role of CPV phases can play an important role in determining the masses of sterile states through an exact seesaw relation and their subsequent contribution to the $\lvert m_{ee}\rvert$. 
Now, the reliability of the obtained values of CPV phases and mixing angles can be verified through the reconciliation of predicted value of the BR of $\mu \rightarrow e \gamma$ and through the extent of non-unitarity arising due to admixture of active-sterile mixing. 
\section{Compatibility with the non-unitary range}
\label{sec:nonunitary-range}
The mixing among active and sterile neutrinos introduces non-unitarity into the mixing matrix 
of the active neutrino sector. As a result, the matrices $V$ and $R$ receive several phenomenological constraints. The quantities of interest are (i) $\left| VV^{\dagger} \right|$, (ii) $\left| V \right|$, and (iii)  $\lvert R \lvert$.  In this section, we calculate the last three quantities 
using the values of various parameters input into the calculation of 
$\lvert m_{ee}  \lvert$ in section (\ref{sec:result}) and compare them with the constraints coming from relevant observations.

Due to the non-unitary in the lepton mixing, the expression for the oscillation probability gets altered (details can be found in Refs. \cite{Antusch:2006vwa, Blennow:2016jkn, Escrihuela:2015wra, Antusch:2009pm, Forero:2021azc}). The modified oscillation probability subsequently triggers the zero-distance effect which in turn imposes stringent limits on the order of deviation from unitarity (maximum up to the the order of $\mathcal{O}(10^{-2})$ \cite{Xing:2007zj, Xing:2009in}). In addition to neutrino oscillation phenomena, the presence of non-unitary impacts on electroweak interactions \cite{Antusch:2016brq, Abada:2007ux} and cLFV processes \cite{Dinh:2012bp, BaBar:2009hkt}. The expressions for various electroweak precision observables and branching ratios of lepton flavor- violating decays of a charged lepton get corrected by the Hermitian combination $\left| VV^{\dagger} \right|$ \cite{Antusch:2006vwa, Das:2017nvm, Antusch:2014woa, Antusch:2016brq}. The range of values of the elements of $\left| VV^{\dagger} \right|$ matrix using the parameter space of our model is given as follows:
\begin{equation}
\lvert VV^{\dagger} \lvert = \begin{pmatrix}
                                  \begin{smallmatrix}
                                 0.999891-0.999999 & 1.52\times 10^{-9} - 6.1.78\times 10^{-5} & 2.06\times 10^{-8} - 9.46\times 10^{-4} \\
..& 0.999717 - 0.999997 & 2.28\times 10^{-7}-1.95\times 10^{-3}\\ .. & .. &    0.999231-0.999993
                                  \end{smallmatrix}
                                  \end{pmatrix}.
\label{eq:vv-nh}
\end{equation}

From Eq. (\ref{eq:vv-nh}), the departure of the diagonal elements of $\left| VV^{\dagger} \right|$ from unity signifies the presence of non-unitary, although it remain minute in scale. However, these minute scale deviations are consistent with the constraint values derived from various electroweak decay processes (e.g., decays of $W$ and $Z$ bosons, universality test in $W$ decays), and the obtained value can be comparable with those given in \cite{Antusch:2006vwa}. Furthermore, the limitations imposed on off-diagonal elements are subjected to the analysis of lepton flavor- violating decays of charged leptons. Specifically, the $(1,2)$ and $(2,1)$ elements receive stringent upper bounds from the $\mu \rightarrow e \gamma$ process, while those of $(2,3)$ and $(3,2)$ are from $\tau \rightarrow \mu \gamma$, and $(1,3)$ and $(3,1)$ are from the $\tau \rightarrow e \gamma$ process.
 As can be seen from Eq. (\ref{eq:vv-nh}), the upper bound on the $(1,2)$ elements comes out to be $\lesssim 10^{-5}$. This limit remains consistent with the imposed restriction on the branching ratio, which is set at BR$(\mu\rightarrow e\gamma)<10^{-13}$ \cite{Das:2017nvm, Antusch:2016brq}. It is important to note that the precision of the branching ratio measurement continues to improve, leading to continuous enhancements in the applied constraints. 
Most of the detectors have been upgraded to improve the search sensitivity by another order of magnitude down to $\sim 10^{-14}$ \cite{baldini2018meg, baldini2021meg} and this sensitivity can be translated into a unitarity constraint on the (1,2) element $< 10^{-6}$ \cite{Antusch:2006vwa}. It can be observed that the representative dataset listed in Tables (\ref{tab:nh-phase}) and (\ref{tab:ih-phase}) satisfies the current constraint value of the BR and the limits on (1,2) elements effectively for the combination of Sets 2 and 3. If the bound on  BR$(\mu\rightarrow e\gamma)$ could be down even further  $< 10^{-15}$ \cite{Cei:2014jtm}, it would put a further constraint on the (1,2) elements  $< 10^{-7}$ \cite{Antusch:2006vwa}, and this sensitivity can be achieved with the combination of Set 1 through 
the parameter space.
 Similarly, the upper bounds on the $(1,3)$ and $(2,3)$ elements, which are attributed to the rare decays of $\tau$-leptons are set at $< 10^{-4}$ and $<10^{-3}$ \cite{Antusch:2006vwa, Das:2017nvm} respectively.

In this context, we have also furnished the comprehensive ranges for the elements of the leptonic mixing matrix as follows:
\begin{equation}
\lvert V \lvert =
\begin{pmatrix}
\substack{0.824991 - 0.825164} & \substack{0.544876 - 0.545001} & \substack{0.149043 - 0.149537}\\
\substack{0.263118 - 0.454206} & \substack{0.479233 -0.604581} & \substack{0.748395 - 0.751832} \\
\substack{0.332458 - 0.500148} & \substack{0.572511 - 0.687958} & \substack{0.631477 - 0.642174} 
\end{pmatrix}.
\label{eq:v-nh}
\end{equation}

The limits presented in Eq. (\ref{eq:v-nh}) match with the constraints derived from various oscillation experiments as reported in
Ref.\cite{Antusch:2006vwa}. Furthermore, with improvement of measurement of mixing matrix elements, the constraints on unitarity violation may be expected from next-generation experiments, as highlighted in Ref. \cite{Ellis:2020hus}.

The degree of unitarity violation is also linked to the $R$ matrix. Its elements $ R_{\alpha k}$ incorporate the active-sterile  mixing. The active-sterile mixing parameters  $ R_{\alpha k}$ are a valuable tool for investigating the potential production of sterile neutrinos at high-energy colliders. As discussed in Ref. \cite{Das:2017nvm} and the references therein, the production cross section is directly proportional to  $\lvert R_{\alpha k}\lvert ^2$. Therefore, through their experimental searches, certain constraints can be imposed on the active-sterile mixing subjected to a varying mass range of sterile neutrinos.  For a deeper understanding, one may refer to Refs.\cite{Bolton:2019pcu, Fang:2021jfv} and the references therein, elucidating a detailed explanation of the distinct constraints on $\lvert R_{\alpha k}\lvert ^2$ corresponding to a specific mass range and the employed experiments to derive these constraints. In line with this, we have explored the overall range of the active-sterile mixing $\lvert R_{\alpha k}\lvert ^2$ using our model parameter space, which is given in Eq. (\ref{eq:r-nh}), whence the upper bound on the elements of $\lvert R \lvert^2$
are found to be $< 10^{-4}$ for both the NH and IH \cite{Das:2017nvm, NA62:2020mcv, T2K:2019jwa}.  
  \begin{equation}
 \lvert R \lvert^2 =
 \begin{pmatrix}
  \substack{2.89\times 10^{-15} - 9.71\times 10^{-7}} &
  \substack{1.50\times 10^{-14} - 3.28\times 10^{-5}}& \substack{3.17\times 10^{-13} - 8.09\times 10^{-5}}\\
\substack{5.76\times 10^{-14} - 2.07\times 10^{-5}}& \substack{2.25\times 10^{-14} - 6.94\times 10^{-6}} & \substack{9.67\times 10^{-11} - 5.59\times 10^{-6}}\\ 
\substack{4.28\times 10^{-13} - 8.12\times 10^{-6}}& \substack{6.32\times 10^{-11} - 1.88\times 10^{-4}}& \substack{4.77\times 10^{-12} - 9.25\times 10^{-4}}
 \end{pmatrix}.
 \label{eq:r-nh}
\end{equation}
\section{cLFV \texorpdfstring{$\mu \rightarrow e \gamma$}~~ decay in 3+3 model}
\label{app:mu-e-gamma}
The massive nature of neutrinos along with violation of their lepton flavor is 
confirmed by the discovery of neutrino oscillation. In this context, the cLFV observables are found to acquire their reasonable value with the addition of neutral leptons \cite{Alonso:2012ji,Abada:2021zcm}.
Among several decay channels, the muon decays are thought to be more prominent due to its largest discovery potential in most of the SM extensions. So, as a promising  probe for cLFV observables like the BR, we find the BR of the $\mu \rightarrow e \gamma$ in the model we have considered and review the parameter space used to calculate $\lvert m_{ee} \rvert$ in section (\ref{sec:result}).
\begin{figure}[htb]
\centerline{\includegraphics[width=3in, height=1.5in]{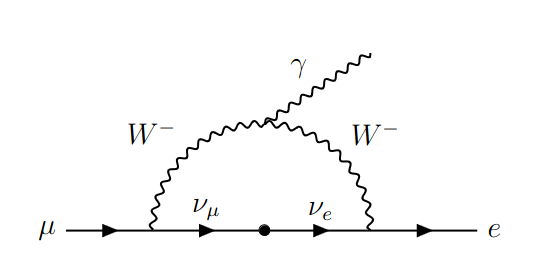}}
\caption{Feynman diagram for $\mu \longrightarrow e\gamma$ decay.}
\label{fig:feyn-meg}
\end{figure}

\begin{figure*}
    \centering
     \begin{subfigure}{0.3\textwidth}
        \centering
        \includegraphics[width=\linewidth]{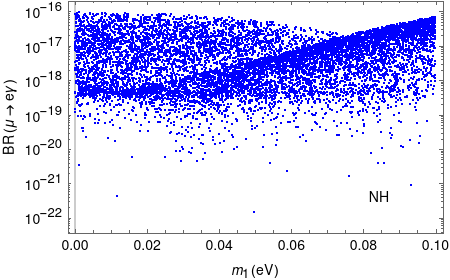}
        \caption{}
        \label{fig:sub19}
    \end{subfigure}
    \hfill
    \begin{subfigure}{0.3\textwidth}
        \centering
        \includegraphics[width=\linewidth]{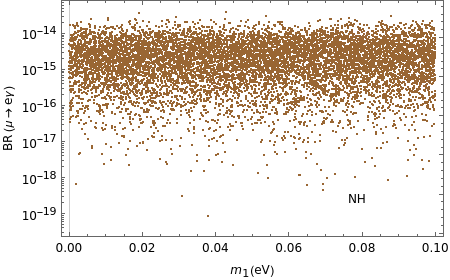}
        \caption{}
        \label{fig:sub20}
    \end{subfigure}
    \hfill
    \begin{subfigure}{0.3\textwidth}
        \centering
        \includegraphics[width=\linewidth]{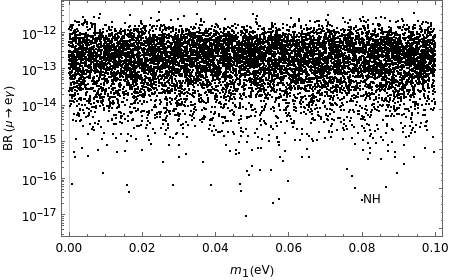}
        \caption{}
        \label{fig:sub21}
    \end{subfigure}

    \medskip

    \begin{subfigure}{0.3\textwidth}
        \centering
        \includegraphics[width=\linewidth]{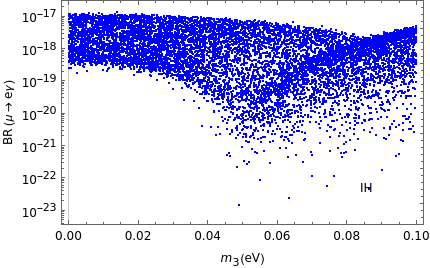}
        \caption{}
        \label{fig:sub22}
    \end{subfigure}
    \hfill
    \begin{subfigure}{0.3\textwidth}
        \centering
        \includegraphics[width=\linewidth]{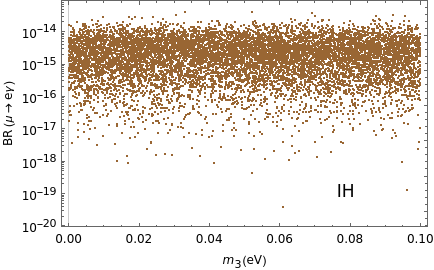}
        \caption{}
        \label{fig:sub23}
    \end{subfigure}
    \hfill
    \begin{subfigure}{0.3\textwidth}
        \centering
        \includegraphics[width=\linewidth]{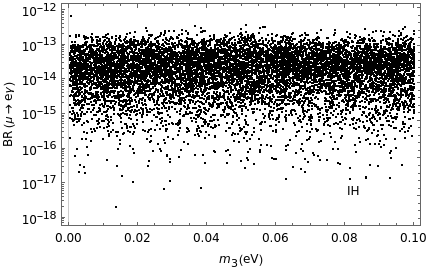}
        \caption{}
        \label{fig:sub24}
    \end{subfigure}
    \caption{ Branching ratio of $\mu \rightarrow e \gamma$ in the parameter space following the same constraints as in the case of $\lvert m_{ee}\lvert$. The upper (lower) panel is for the NH (IH) of active neutrinos. The left, middle, and right columns corresponds to the combination 
of active-sterile mixing angles of Sets 1, 2, and 3, respectively.}
    \label{fig:br-meg}
\end{figure*}

From the loop diagram shown in Fig.(\ref{fig:feyn-meg}), the branching ratio of this process can be simply approximated by considering that it is essentially given by the product of three factors: (i) the usual muon decay, (ii) an electromagnetic vertex for photon emission, (iii) the neutrino mixing.
However, calculations of BR($\mu \rightarrow e \gamma$) in the SM limit can be found in \cite{Hambye:2013jsa, Ardu:2022sbt, Galli:2019xop}, and the same with an SM extension via N sterile states has been furnished in \cite{Abada:2021zcm, Alonso:2012ji, Ilakovac:1994kj}. Likewise, in the model we have considered, the expression for BR($\mu \rightarrow e \gamma$) gets revamped to
\begin{equation}
BR(\mu \rightarrow e\gamma) = \frac{\alpha_w^3 s_w^2}{256 \pi^2}
\frac{m_\mu^4}{M_W^4}\frac{m_\mu}{\Gamma_\mu}
\left|G_\gamma^{\mu e}\right|^2,
\label{eq:br-meg}
\end{equation}
where $G_\gamma^{\mu e}$ is the form factor and is given by 
\begin{equation}
G_\gamma^{\mu e}=\sum_{i=1}^{3+3}\mathcal{U}^*_{\mu i}\mathcal{U}_{ei}G_\gamma(x_i).
\label{eq:formfactor}
\end{equation}
Here, $\alpha_w$ is the weak coupling, $s_w =\sin \theta_w$; $\theta_w$ is the weak mixing angle, $m_\mu$ and $\Gamma_\mu$ are the mass and total decay width of the muon, respectively, and $M_W$ is the mass of W boson. The numerical values of these factors can be found in Refs. \cite{ParticleDataGroup:1994kdp, Alonso:2012ji}.
With $x_i=m_i^2/M_W^2$ ($m_i$ corresponds to the respective masses of active and added sterile neutrinos), the loop function is defined as (taken from \cite{Abada:2021zcm, Ilakovac:1994kj}):
\begin{equation}
G_\gamma(x_i)= - \frac{x_i(2x_i^2 + 5x_i -1)}{4(1-x_i)^3}-\frac{3x_i^3}{2(1-x_i)^4} \log{x_i}.
\label{eq:loop-func}
\end{equation}
Now we have calculated the BR using Eq.(\ref{eq:br-meg}) and with those phases and mixing angles scrutinized for $0\nu \beta \beta$ decay and obtained the results as shown in Fig.(\ref{fig:br-meg}).

 The ranges for the BR obtained from numerical calculation are found to match with both the current experimental sensitivity, i.e., $10^{-13} - 10^{-14}$ \cite{baldini2018meg, baldini2021meg, MEG:2011naj}, as well as the future sensitivity $< 10^{-15}$ \cite{Cei:2014jtm}. From Fig.(\ref{fig:br-meg}), the BR  seems to be insensitive to $m_{\rm lightest}$ except for Set 1 for the IH. But the CPV phases $\phi_{1i}$ and active-sterile mixing angles seem to play an important role in predicting  ${\textrm{BR}}({\mu \rightarrow e \gamma})$ in the experimental sensitive region. For the purpose of illustration, we have identified a small set of representative points which includes the phase, active-sterile mixing angle values, and the corresponding BR and $\lvert m_{ee}\lvert$ in Table (\ref{tab:nh-phase}) for the NH and Table (\ref{tab:ih-phase}) for the IH.

\begin{table}[hbt]
    \centering
    \caption{Some of the benchmark points used to calculate masses of the
	sterile states, effective mass, and the BR($\mu\longrightarrow e \gamma$) for the NH of active neutrinos.}
    \label{tab:nh-phase}
    \begin{tabular}{ccc}
        \toprule
        \multicolumn{1}{c}{Phases} & \multicolumn{1}{c}{Mixing angles} & \multicolumn{1}{c}{Observables} \\
        \midrule
        \midrule
     $\phi_{12}=6.0658$, $\phi_{13}=1.4538$  & $\theta_{14}=1.96\times 10^{-5}, \theta_{24}=2.42\times 10^{-3}$, $\theta_{34}=2.04\times 10^{-2}$ & $M_1=0.11~ {\rm TeV}$, $M_2=8.33~ {\rm GeV}$\\
       $\phi_{14}=4.0617$, $\phi_{15}=0.6455$&$\theta_{15}=8.58\times 10^{-4}$, $\theta_{25}=8.74\times 10^{-4}, \theta_{35}=7.50\times 10^{-2}$ & $M_3=0.65~ {\rm GeV}$, $\lvert m_{ee}\lvert=0.011 \;{\rm eV}$\\
		$\phi_{16}=1.7991$&  $\theta_{16}=1.88\times 10^{-4}, \theta_{26}=5.16\times 10^{-3}$, $\theta_{36}=4.15\times 10^{-2}$& ${\rm BR}=3.17\times10^{-13}$\\
		\hline
		 $\phi_{12}=0.4578$, $\phi_{13}=6.1057$  & $\theta_{14}=8.97\times 10^{-5}, \theta_{24}=6.94\times 10^{-3}$, $\theta_{34}=7.40\times 10^{-2}$ & $M_1=0.68~ {\rm TeV}$, $M_2=0.23~ {\rm TeV}$\\
       $\phi_{14}=5.0806$, $\phi_{15}=5.4469$&$\theta_{15}=1.25\times 10^{-4}$, $\theta_{25}=2.63\times 10^{-3}, \theta_{35}=1.83\times 10^{-2}$ & $M_3=0.67~ {\rm TeV}$, $\lvert m_{ee}\lvert=0.0086 \;{\rm eV}$\\
		$\phi_{16}=4.7143$&  $\theta_{16}=7.40\times 10^{-5}, \theta_{26}=3.57\times 10^{-3}$, $\theta_{36}=7.41\times 10^{-2}$& ${\rm BR}=2.43\times10^{-14}$\\
		\hline
    $\phi_{12}=5.7419$, $\phi_{13}=3.2117$  & $\theta_{14}=1.30\times 10^{-5}, \theta_{24}=9.07\times 10^{-4}$, $\theta_{34}=1.32\times 10^{-3}$ & $M_1=0.64~ {\rm TeV}$, $M_2=82.2~ {\rm TeV}$\\
       $\phi_{14}=0.9568$, $\phi_{15}=5.0211$&$\theta_{15}=1.07\times 10^{-5}$, $\theta_{25}=3.49\times 10^{-4}, \theta_{35}=2.43\times 10^{-3}$ & $M_3=2.54~ {\rm TeV}$, $\lvert m_{ee}\lvert=0.084 \;eV$\\
$\phi_{16}=0.9284$&  $\theta_{16}=2.53\times 10^{-6}, \theta_{26}=2.83\times 10^{-4}$, $\theta_{36}=1.61\times 10^{-3}$& ${\rm BR}=9.15\times10^{-16}$\\
        \bottomrule

    \end{tabular}
\end{table}

\begin{table}[hbt]
    \centering
    \caption{Some of the benchmark points used to calculate masses of the
	sterile states, effective mass and the BR($\mu\longrightarrow e \gamma$) for the IH of active neutrinos.}
    \label{tab:ih-phase}
    \begin{tabular}{ccc}
        \toprule
        \multicolumn{1}{c}{Phases} & \multicolumn{1}{c}{Mixing angles} & \multicolumn{1}{c}{Observables} \\
        \midrule
        \midrule
      $\phi_{12}=4.8466$, $\phi_{13}=3.1790$  & $\theta_{14}=1.45\times 10^{-5}, \theta_{24}=6.68\times 10^{-4}$, $\theta_{34}=1.52\times 10^{-3}$ & $M_1=11~ {\rm GeV}$, $M_2=0.3~ {\rm GeV}$\\
       $\phi_{14}=3.7868$, $\phi_{15}=0.1489$&$\theta_{15}=1.56\times 10^{-4}$, $\theta_{25}=1.48\times 10^{-3}, \theta_{35}=7.37\times 10^{-3}$ & $M_3=0.8~ {\rm TeV}$, $\lvert m_{ee}\lvert=0.046 \;{\rm eV}$\\
		$\phi_{16}=0.0899$&  $\theta_{16}=1.44\times 10^{-5}, \theta_{26}=6.89\times 10^{-3}$, $\theta_{36}=7.55\times 10^{-2}$& ${\rm BR}=1.82\times10^{-13}$\\
		\hline
		$\phi_{12}=1.6683$, $\phi_{13}=4.9542$  & $\theta_{14}=6.11\times 10^{-4}, \theta_{24}=1.15\times 10^{-3}$, $\theta_{34}=4.56\times 10^{-2}$ & $M_1=0.93~ {\rm GeV}$, $M_2=60~ {\rm GeV}$\\
       $\phi_{14}=4.1378$, $\phi_{15}=6.1501$&$\theta_{15}=5.98\times 10^{-5}$, $\theta_{25}=5.05\times 10^{-3}, \theta_{35}=1.45\times 10^{-2}$ & $M_3=10~ {\rm GeV}$, $\lvert m_{ee}\lvert=0.026 \;{\rm eV}$\\
		$\phi_{16}=0.8690$&  $\theta_{16}=2.01\times 10^{-4}, \theta_{26}=9.83\times 10^{-3}$, $\theta_{36}=3.20\times 10^{-2}$& ${\rm BR}=7.42\times10^{-14}$\\
		\hline
   $\phi_{12}=4.8071$, $\phi_{13}=3.0606$  & $\theta_{14}=1.30\times 10^{-6}, \theta_{24}=7.61\times 10^{-5}$, $\theta_{34}=1.51\times 10^{-3}$ & $M_1=16.01~ {\rm TeV}$, $M_2=0.53~ {\rm TeV}$\\
       $\phi_{14}=2.6423$, $\phi_{15}=3.3156$&$\theta_{15}=5.50\times 10^{-6}$, $\theta_{25}=1.17\times 10^{-4}, \theta_{35}=1.46\times 10^{-3}$ & $M_3=34~ {\rm TeV}$, $\lvert m_{ee}\lvert=0.12 \;eV$\\
$\phi_{16}=2.8832$&  $\theta_{16}=5.68\times 10^{-5}, \theta_{26}=1.03\times 10^{-4}$, $\theta_{36}=5.22\times 10^{-3}$& ${\rm BR}=1.89\times10^{-17}$\\
        \bottomrule

    \end{tabular}
\end{table}

\section{Conclusion}
\label{sec:conclusion}
The present and future results from neutrino oscillation data on neutrino mixing and mass hierarchy, along with constraints on absolute neutrino mass from cosmology, can predict the contribution of light active neutrinos to $0\nu\beta\beta$ decay. If this contribution aligns with future $0\nu\beta\beta$ observations, it would establish that light active neutrinos dominate the $0\nu\beta\beta$ decay process. Any other observation detected in the searches for $0\nu\beta\beta$ decay would call for new physics contributions.

In this work, we have studied the combined contributions of active and sterile neutrinos to speculate on the effective electron neutrino mass $\lvert m_{ee}\rvert$ in $0\nu \beta \beta$ decay. In summary, we explored the role of sterile neutrinos in the $0\nu\beta\beta$ decay process within an extension of the SM. We obtained a $6\times6$ leptonic mixing matrix encompassing active-sterile mixing and CPV phases. Using the exact seesaw relation, we derived three analytical expressions for the masses of the added sterile states $M_k$ in terms of active-sterile mixing angles, CPV phases, and the lightest active neutrino mass. These expressions impose certain constraints on the selected CPV phases.
 Varying the phases within $0$ and $2\pi$ and taking three benchmark sets
of active-sterile mixing angles, we find the possible mass ranges of the sterile states as a function of the lightest mass of the active neutrinos.
The parameter space was explored keeping the prospects of direct searches of   
sterile neutrinos under consideration.

For different benchmark sets, the masses were found to range from $\sim$ keV to TeV. In this scenario, the masses of the sterile states, derived from the lightest mass of active neutrinos and active-active and active-sterile mixing angles and phases, can be all light, all heavy, or a mixture of both compared to the active states. This variation is attributed to the interplay of the CPV phases of the $6\times 6$ mixing matrix. Using the constrained parameter space,
the combined contribution of active-sterile neutrino to $\lvert m_{ee}\rvert$
was calculated to gauge the implication of direct searches from laboratory and 
cosmology on the $0\nu\beta\beta$ decay process. This analysis presents a constrained scenario of searches of sterile neutrinos by considering that 
light neutrinos are generated via a type-I seesaw mechanism. The consequences 
of an exact seesaw relation present a direct link between active and sterile sectors.

The numerical values obtained for $\lvert m_{ee}\rvert$ from the combined contribution of both active and sterile neutrinos can vary from as low as $\mathcal{O}(10^{-4})$ to saturating the present experimental limit. The parameter space of $\lvert m_{ee}\lvert$ versus $m_{\rm lightest}$ 
is sensitive to the combination of active-sterile mixing angles.
Also, the results from the combined contribution are found to saturate the experimental bound for lower values of the lightest neutrino mass as compared to the standard three-neutrino picture for the NH. 
Future cosmological observations are expected to probe the sum of active neutrino masses in the range $0.03 - 0.06$ eV \cite{Amendola:2016saw,SPT-3G:2019sok,CMB-S4:2022ght}. The future ton-scale experiments like CUPID \cite{CUPID:2019imh,Armengaud:2019loe}, LEGEND \cite{LEGEND:2017cdu}, and nEXO \cite{nEXO:2017nam,nEXO:2018ylp} are going to probe $\lvert m_{ee} \rvert \le 0.01$ eV. 
With the projected sensitivity of the sum of active neutrino masses and if 
$0\nu\beta \beta$ decay is registered with $\lvert m_{ee}\rvert \lesssim 0.01$ eV,
the standard three neutrino contribution to $\lvert m_{ee}\rvert$ will be in conflict with the observations. But the combined contribution will still offer some parameter
space to study new physics contributions to the $0\nu\beta\beta$ decay process.
In this case, there will be some parameter space available to probe for new physics contributions dominating the $0\nu\beta\beta$ decay.  The reliability of our results was also verified by calculating the non-unitarity effect and the branching ratio of $\mu \rightarrow e \gamma$, a prominent cLFV process, in this framework.

\appendix
\section{Role of nuclear matrix elements in \texorpdfstring{$0\nu\beta\beta$}{} decay process}
\label{app:nme}
In this section, we outline the role of nuclear matrix elements
in their contribution to $\lvert m_{ee} \rvert$, as has been studied in Ref. \cite{Blennow:2010th},
for different cases of mass hierarchy among the sterile
neutrinos. From Eq.(\ref{eq:timeperiod}), the amplitude of $0\nu\beta\beta$ decay in the presence of active and sterile states can be expressed as 
\begin{equation}
 A \propto \sum_i m_i \mathcal{U}^2_{ei}M^{0\nu\beta\beta}(m_i) + 
 \sum_k M_k \mathcal{U}_{ek}^2 M^{0\nu\beta\beta}(M_k),
 \label{eq:ampli}
\end{equation}
where $m_i$ and $M_k$ are the masses of the propagating active and sterile neutrinos, respectively, and $M^{0\nu\beta\beta}(m_i)$ and \\ $M^{0\nu\beta\beta}(M_k)$ are the corresponding
nuclear matrix elements. Using the parameterization given in Eq.(\ref{eq:curl-U}), the amplitude can further be expressed as
\begin{equation}
 A \propto \sum_i m_i V^2_{ei}M^{0\nu\beta\beta}(m_i) + 
 \sum_k M_k R_{ek}^2 M^{0\nu\beta\beta}(M_k).
 \label{eq:ampli-1}
\end{equation}
The propagator in the $0\nu\beta\beta$ decay process, $1/(p^2 -m^2)$ (where $p^2 \sim -(100~ {\rm MeV})^2$) decides the 
behavior of nuclear matrix elements.
The nuclear matrix elements are constant for $M< 100$ MeV and decrease as 
$M^{-2}$ for $M> 100$ MeV. The transition around $100$ MeV is smooth, and no
significantly new physics takes place at this regime. So, depending upon the masses of the  sterile states, the amplitude of $0\nu\beta\beta$ decay can be implicated with the exact seesaw relation in a considerable way as follows:
\begin{itemize}
 \item {\bf All sterile states are in the light regime.}\\
 In this case, using Eq.(\ref{eq:ee-element}),
 \begin{equation}
\sum_{i=1}^{\rm light} m_i V^2_{ei} + \sum_{k=1}^{\rm light} M_k R^2_{ek} = 0,
\label{eq:ee-element-light}
\end{equation}
Thus,
 \begin{eqnarray}
 \nonumber
 A &\propto&  \sum^{\rm light}_i m_i \mathcal{U}^2_{ei}M^{0\nu\beta\beta}(m_i) + 
 \sum^{\rm light}_K M_k \mathcal{U}_{ek}^2 M^{0\nu\beta\beta}(M_k)\\
 &\approx& - \sum^{\rm light}_k M_k \mathcal{U}_{ek}^2\left( M^{0\nu\beta\beta}(0)
 -  M^{0\nu\beta\beta}(M_k) \right)
 \label{eq:ampli-light}
\end{eqnarray}
For $M_k < 1$ MeV, the $( M^{0\nu\beta\beta}(0)
 -  M^{0\nu\beta\beta}(M_k)$ suppression becomes almost $10^{-3}$, so that the 
 $0\nu\beta\beta$ decay rate should be suppressed by six orders of magnitude. The 
 decay rate is even more suppressed for lower values of $M_k$. So, for this regime
 of masses of sterile neutrinos, $0\nu\beta\beta$ decay becomes experimentally 
 inaccessible. However, a detailed study of nuclear matrix elements, especially in the light mass regime, and its impact on the $0\nu\beta\beta$ decay amplitude has been carried out in Refs.\cite{Dekens:2023iyc, Dekens:2020ttz}, where a reduced level of suppression rate is emphasized.
 \item {\bf All sterile states are in the heavy regime.}\\
 Proceeding as above,
 \begin{equation}
\sum_{i=1}^{\rm light} m_i V^2_{ei} + \sum_{k=1}^{\rm heavy} M_k R^2_{ek} = 0.
\label{eq:ee-element-heavy}
\end{equation}
 In this case,
 \begin{eqnarray}
 \nonumber
 A &\propto&  \sum^{\rm light}_i m_i \mathcal{U}^2_{ei}M^{0\nu\beta\beta}(m_i) + 
 \sum^{\rm heavy}_k M_k \mathcal{U}_{ek}^2 M^{0\nu\beta\beta}(M_k)\\
 &\approx& - \sum^{\rm heavy}_k M_k \mathcal{U}_{ek}^2\left( M^{0\nu\beta\beta}(0)
 -  M^{0\nu\beta\beta}(M_k) \right)\\ \nonumber
 &\approx&  - \sum^{\rm heavy}_k M_k \mathcal{U}_{ek}^2 M^{0\nu\beta\beta}(0) =
 \sum^{\rm light}_i m_i \mathcal{U}^2_{ei}M^{0\nu\beta\beta}(0)
 \label{eq:ampli-heavy}
\end{eqnarray}
In this case, the contribution from the light active neutrinos dominates the 
$0\nu\beta\beta$ decay rate.
 \item {\bf  Sterile states in the light and heavy regime.}\\
Similarly as above, 
\begin{equation}
 \sum_{i=1}^{\rm light} m_i V^2_{ei} + \sum_{k=1}^{\rm light} M_k R^2_{ek} 
+ \sum_{k=1}^{\rm heavy} M_k R^2_{ek} = 0,
\label{eq:ee-element-heavy-light}
\end{equation}
Since for the sterile states in the heavy regime, the nuclear matrix elements
receive extra suppression to their contribution to the $0\nu\beta\beta$ decay,
the leading terms arise from light states
\begin{eqnarray}
 \nonumber
 A &\propto&  \sum^{\rm light}_i m_i \mathcal{U}^2_{ei}M^{0\nu\beta\beta}(m_i) + 
 \sum^{\rm light}_k M_k \mathcal{U}_{ek}^2 M^{0\nu\beta\beta}(M_k)\\
 &\approx& - \sum^{\rm heavy}_k M_k \mathcal{U}_{ek}^2 M^{0\nu\beta\beta}(0)
  \label{eq:ampli-light-heavy}
\end{eqnarray}
\end{itemize}
This scenario is phenomenologically the most promising. By following Eq.(\ref{eq:ampli-light-heavy}), 
the result could be interpreted as 
\begin{equation}
 0.11 ~{\rm eV} < \lvert \sum^{\rm heavy}_k M_k \mathcal{U}_{ek}^2 \rvert < 0.56~ {\rm eV}.
\end{equation}
 This large contribution to $0\nu\beta\beta$ decay would not
be in conflict with the neutrino masses if the light and heavy sterile neutrino
contributions cancel each other in Eq.(\ref{eq:ee-element-heavy-light}).

\end{document}